# Identifying spatial interdependence in panel data with large N and small T

Deborah Gefang

School of Business, Leicester University

Stephen G. Hall

School of Business, Leicester University, Bank of Greece,
and University of Pretoria

George S. Tavlas

Bank of Greece and the Hoover Institution, Stanford University

## Abstract

This paper develops a simple two-stage variational Bayesian algorithm to estimate panel spatial autoregressive models, where $N$, the number of cross-sectional units, is much larger than $T$, the number of time periods without restricting the spatial effects using a predetermined weighting matrix. We use Dirichlet-Laplace priors for variable selection and parameter shrinkage. Without imposing any a priori structures on the spatial linkages between variables, we let the data speak for themselves. Extensive Monte Carlo studies show that our method is super-fast and our estimated spatial weights matrices strongly resemble the true spatial weights matrices. As an illustration, we investigate the spatial interdependence of European Union regional gross value added growth rates. In addition to a clear pattern of predominant country clusters, we have uncovered a number of important between-country spatial linkages which are yet to be documented in the literature. This new procedure for estimating spatial effects is of particular relevance for researchers and policy makers alike.

1. **Introduction**

Since the seminal paper by Cliff and Ord (1973), spatial autoregressive (SAR) models have been widely used in the literature to investigate the cross-sectional interdependencies between variables (e.g., Anselin 1988, Baltagi et al. 2003 2013, Lee and Yu 2010, LeSage and Pace 2018, to mention a few). In a SAR model, the spatial structure between units is captured by the $N \times N$ spatial weights matrix with zero diagonal entries. Since it is often challenging to estimate the $N^2 - N$ off-diagonal elements in a spatial weights matrix, most of the available studies elicit the matrices a priori with references to theories and conventions.

As summarized in Fingleton and Arbia (2008), a major criticism of SAR models is that their inferences such as the spillover effects are sensitive to how the spatial weights matrix is specified. There are two major drawbacks of imposing the predetermined spatial weights matrices to a model without letting the data speak. To start with, economic theories and conventions are not able to provide researchers with precice spatial weights matrix entries. In addition, for the same data, different theories and conventions may imply very different spatial linkage relationships, which may even contradict each other. Studies, such as those by Kelejian (2008) and Kelejian and Piras (2011, 2016), have developed various advanced testing methods to select the 'true spatial weights matrix' among a number of plausible candidates. However, the success of the selection procedure depends on the 'true spatial weights matrix', which is hard to pin down, being included in the candidates pool in the first place. Hence, it is important to work with models in which the spatial weights matrices are not set a priori.

Early efforts to estimate the spatial weights matrices usually relied on imposing less restrictive assumptions on the spatial structure (e.g., Pinkse et al. 2002, Bhattacharjee and Butler 2013, Bailey et al. 2016). In recent years, developments in various variable selection and parameter shrinkage techniques suitable for estimating models where the number of parameters is larger than the number of observations have equipped researchers with tools to estimate the spatial weights matrices directly from panel data. Using these developments, Ahrens and Bhattacharjee (2015) introduce a two-step Lasso estimator to uncover an unrestricted spatial weights matrix; Lam and Souza (2020) propose to combine prior knowledge on spatial interdependence with data information captured by a Lasso estimator; Gefang et al. (2023) use a two-stage variational Bayesian approach with Dirichlet-Lasso (D-L) prior to estimate the unrestricted spatial weights matrix; and Krisztin and Piribauer (2023) introduce a Bayesian approach to estimate a spatial weights matrix with its off-diagonal elements being either 0 or 1 before being row-standardised.

In comparison with the methods proposed in Ahrens and Bhattacharjee (2015) and Gefang et al. (2023), methods of Lam and Souza (2020) and Krisztin and Piribauer (2023) can be less appealing as the former only works when the spatial linkages conjectured by the researchers are not too different from the true one, while the latter rules out the possibility of any negative elements in a spatial weights matrix. In addition, both Lam and Souza (2020) and Krisztin and Piribauer (2023) assume the same spatial parameters for all units, while Ahrens and Bhattacharjee (2015) and Gefang et al. (2023) do not impose such restrictions.

Following Ahrens and Bhattacharjee (2015) and Gefang et al. (2023), this paper focuses on spatial panel data models with an unrestricted spatial weights matrix in order to let the data speak. More specifically, we introduce a simple two-stage variational Bayesian (VB) approach to estimating spatial panel data model where the ratio of $N$ to $T$ is much larger than those that can be dealt with in Ahrens and Bhattacharjee (2015) and Gefang et al. (2023). We use Dirichlet-Laplace priors for variable selection and parameter shrinkage. Despite the caveat that global-local-shrinkage priors tend to shrink all the non-diagonal spatial weights entries to zero, but never to zero, when $N$ is much larger than $T$, our estimated spatial weights matrices turn out to be highly similar to their true counterparts. This feature is of particular importance to researchers. Note that Lesage and Pace (2014) have shown that when two spatial weights matrices are similar, the spatial effects are not that sensitive to the spatial matrix used. Their findings are corroborated in our Monte Carlo excersises.

Extensive Monte Carlo exercises show that our simple two-stage VB approach works well. Keeping $T$ as small as 20 , we are able to uncover the spatial relationships for $N$ as large as 500 (which implies 249,500 spatial parameters to be estimated). Noting the limitation of traditional spatial weights matrices that they only contain non-negative entries, we specifically allow for both positive and negative spatial weights in the simulations. Monte Carlo results show that our estimated spatial weights matrices are similar to the true ones. Interestingly, keeping $T$ unchanged, while the simple correlations[1] between the estimated and true spatial weights matrices decreases while $N$ increases, the structural similarity index measure (SSIM)[2] between those two matrices goes up, indicating when noises are taken out, large $N$ is actually helpful in uncovering the true spatial weights matrices.

We also compare and contrast between the true direct effects and indirect effects, which are used to measure the spillovers between variables. As we briefly mentioned above, our Monta Carlo results are very encouraging in the sense that the estimated direct effects are almost perfectly positively correlated with true direct effects, and the estimated indirect effects are highly correlated

---

[1] Simple correlation between two matrices $A$ and $B$ is

$$corr2 = \frac{\sum_n \sum_m (A_{mn} - \bar{A})(B_{mn} - \bar{B})}{\sqrt{(\sum_n \sum_m (A_{mn} - \bar{A})^2)(\sum_n \sum_m (A_{mn} - \bar{A})^2)}}.$$

[2] Details about SSIM can be found in Zhou et al. (2004).

with the true indirect effects. Despite the drawbacks that, when $N$ becomes larger, the magnitude of the estimated direct effects tend to be larger than those of the true ones, and the magnitudes of the estimated indirect effects tend to be smaller than their true values, at the minimum, we can use the estimated effects matrices to identify the spillover patterns between variables with confidence.

For emprical application we examine the spatial interconections and spillovers between European Union (EU) regional gross value added (GVA) growths. Due to its important economic and policy implications, spatial interdependence of regional economies has been intensively investigated since early 2000s. A majority of the studies set a priori spatial weights matrices based on geographical distances and social economic proximities (e.g., Arbia et al. 2010, Basile et al. 2014, Crespo et al. 2014, and Piribauer 2016, to mention a few). Estimating spatial weights matrix from the data, an interesting study by Piribauer et al. (2023) finds significant country clusters and marked divide between regions in the Western and Northern Europe and those in Southern and Eastern Europe. The spatial weights matrix used in that study, however, can nevertheless be considered as quite restrictive. To start with, the authors assume all the non-zero entries in the same row of the spatial weights matrix are of the same value, implying all regions spatially related to an individual region shall have the same spillover effects on that region, which can be unrealistic. Furthermore, the entries of their spatial weights matrix are all non-negative, making it impossible to account for phenomenon where a region's developments hollow out its spatially related neighbours.

Using an unrestricted spatial weights matrix, we find pronouced country clusters dominating other spatial connections, as in Piribauer et al. (2023). But the country groupings we uncovered are more complicated than previously known. To start with, we find that a small number of regions are associated with negative spacial weights , implying their economic growth might be attracting resources from other regions, at the latters' expense. In addition, we find that in comparison with other major EU economies, the German economy is much less linked with other country's economies. Moreover, instead of the well documented divides, such as the north-south and west-east divides, between country groups, we find regions in countries such as France, Spain, Portugal and Italy share clear spatial linkages. We also find a number of other interesting spatial interdenpendencies which have potential important policy implications. For example, the Greek economy is closely spatially affected by Spain's, but much less so by other countries'.

The rest of the paper is organised as follows. Section 2 describes the unrestricted panel SAR model and the two-stage VB estimation approach. Section 3 conducts Monte Carlo excercises to examine the performance of the two-stage VB approach. Section 4 uses an application on EU GVA data to show how our modeling framework and estimation technique can shed important new light on a topic that has been extensively investigated in the literature. Section 5 concludes.

2. **Panel SAR model with unrestricted spatial weights matrix**

A traditional panel SAR model usuallly takes the following form:

$$\underbrace{\begin{bmatrix} y_{1t} \\ y_{2t} \\ \vdots \\ y_{Nt} \end{bmatrix} = \rho \begin{bmatrix} 0 & W_{12} & \cdots & W_{1N} \\ W_{21} & 0 & \cdots & W_{2N} \\ \vdots & \vdots & \cdots & \vdots \\ W_{N1} & W_{N2} & \cdots & 0 \end{bmatrix} \begin{bmatrix} y_{1t} \\ y_{2t} \\ \vdots \\ y_{Nt} \end{bmatrix} + \begin{bmatrix} x_{1t,1} & \cdots & x_{1t,k} \\ x_{2t,1} & \cdots & x_{2t,k} \\ \vdots & \cdots & \vdots \\ x_{Nt,1} & \cdots & x_{Nt,k} \end{bmatrix} \begin{bmatrix} \theta_1 \\ \vdots \\ \theta_N \end{bmatrix} + \begin{bmatrix} \varepsilon_{1t} \\ \varepsilon_{2t} \\ \vdots \\ \varepsilon_{Nt} \end{bmatrix}}_{y_t = \rho W y_t + x_t \theta + \varepsilon_t} \quad (1)$$

where $y_t$ is an $N \times 1$ vector of dependent variables, $W$ is the known row-standardised $N \times N$ spatial weights matrix with zero diagonal entries, $x_t$ is the $N \times k$ matrix of exogenous explanatory variables which may include intercepts, $\rho$ is the spatial parameter, which along with coefficients $\theta$, are to be estimated from the data, and $\varepsilon_t \overset{iid}{\sim} N(0, \sigma^2 I_N)$.[3]

The major drawbacks of model (1) are threefold. First, and most importantly, in (1), *W* is exogenously pre-imposed. Although the spatial parameter $\rho$ can be estimated from the data, if *W* is misspesified it is impossible to pin down the true spatial relationship between $y_{it}$ and $y_{jt}$, and $\rho$ will not be consistently estimated. Second, even with a sensible *W*, (1) is too restrictive in the sense that $\rho$ can only reflect the average level of dependence across all the spatial related units, while it is often more desirable to evaluate how $y_{it}$ and $y_{jt}$ is each influenced by their spatial related units, respectively. Last, but not the least, in model (1), the off-diagonal elements in *W* are usually assumed to be nonnegative (and often symmetric), which rules out the possibility of any negative spillovers such as those caused by crowding-out effects.

---

[3] We assume $\rho \in (-1,1)$ to ensure that the process is stationary.

By contrast, the spatial panel data model of Ahrens and Bhattacharjee (2015) and Gefang et al. (2023) takes the following form:

$$\begin{bmatrix} y_{1t} \\ y_{2t} \\ \vdots \\ y_{Nt} \end{bmatrix} = \begin{bmatrix} 0 & W_{12}^* & \cdots & W_{1N}^* \\ W_{21}^* & 0 & \cdots & W_{2N}^* \\ \vdots & \vdots & \cdots & \vdots \\ W_{N1}^* & W_{N2}^* & \cdots & 0 \end{bmatrix} \begin{bmatrix} y_{1t} \\ y_{2t} \\ \vdots \\ y_{Nt} \end{bmatrix} + \begin{bmatrix} x_{1t,1} & \cdots & x_{1t,k} & 0 & \cdots & 0 & \cdots & 0 & \cdots & 0 \\ 0 & \cdots & 0 & x_{2t,1} & \cdots & x_{2t,k} & \cdots & 0 & \cdots & 0 \\ \vdots & \cdots & \vdots & \vdots & \cdots & \vdots & \cdots & \vdots & \cdots & \vdots \\ 0 & \cdots & 0 & 0 & \cdots & 0 & \cdots & x_{Nt,1} & \cdots & x_{Nt,k} \end{bmatrix} \begin{bmatrix} \theta_{1,1} \\ \vdots \\ \theta_{1,k} \\ \theta_{2,1} \\ \vdots \\ \theta_{2,k} \\ \vdots \\ \theta_{N,1} \\ \vdots \\ \theta_{N,k} \end{bmatrix} + \begin{bmatrix} \varepsilon_{1t} \\ \varepsilon_{2t} \\ \vdots \\ \varepsilon_{Nt} \end{bmatrix} \quad (2)$$

$$\underbrace{\phantom{XX}}_{y_t = W^* y_t + x_t^* \theta^* + \varepsilon_t^*}$$

Unlike $W$ in model (1), which is pre-imposed, the off-diagonal elements of the $N \times N$ spatial weights matrix $W^*$ (again with zero diagonal entries) in model (2) are estimated from the data.[4] If we row-standardise $W^*$, the sum of each of its row can be treated as a spatial parameter similar to $\rho$ shown in model (1). Unlike in model (1) where $\rho$ remains the same for all units, here we have different spatial parameters for different unit, hence $\rho_i$ is allowed to be different from $\rho_j$, if $i \neq j$. Finally, model (2) allows for different coefficients for $x_{it}$ and $x_{jt}$ and

$$\varepsilon_t^* \overset{iid}{\sim} N\left( \begin{bmatrix} 0 \\ 0 \\ \vdots \\ 0 \end{bmatrix}, \begin{bmatrix} \sigma_1^2 & 0 & \cdots & 0 \\ 0 & \sigma_2^2 & \cdots & 0 \\ \vdots & \vdots & \cdots & \vdots \\ 0 & 0 & \cdots & \sigma_N^2 \end{bmatrix} \right) \text{ to allow for more flexibility.}$$

---

[4] We assume the infinity-norm of $W^*$ to be less than 1 to ensure stationarity, where the infinity-norm is defined as

$$\|W^*\|_\infty = \max_{1 \leq i \leq N} \left( \sum_{n=1}^N |W_{in}^*| \right).$$

## 2.1 Bayesian Estimation Techniques

As Piribauer et al (2023) rightly observed, the methods of Ahrens and Bhattacharjee (2015) and Gefang et al (2023) are not suitable for panel data where $N > 200$ while $T$ remains rather small. Since panel data are often featured by large $N$ and small $T$, it is important to develop an approach to estimating model (2) that can surmount the above mentioned limits. Keeping that in mind, in what follows we propose the following simple two-stage VB estimation approach.

First, using all the exogenous (or predetermined) $x_t$ as instrumental variables to calculate the predicted values of $y_{nt}$ for $n = 1,...,N$; Second, estimating the equation where $y_{it}$ is the dependent variable, by substituting $y_{jt}$s, for $j \in \{1,...,N\}$ and $j \neq i$, on the right-hand-side of the equation, by their predicted values derived in the first stage. In essence, those two stages are similar to the two-stage least squares (2LSL) in which the parameters can be estimated equation by equation. It is attractive because the estimations can be performed in parallel for each stage.

Our method is Bayesian. More specifically, we use D-L prior of Bhattacharya et al. (2015) for variable selection and parameter shrinkage. As with other popular global–local shrinkage priors (e.g., Poison and Scott 2012a,b, Carvalho et al. 2010), the computing cost of D-L priros is relatively low, hence suitable for estimating models with a large number of parameters. Moreover, with D-L prior, the entire posterior distribution concentrates at the optimal rate (Bhattacharya et al, 2015). Finally, studies such as Zhang and Bondell (2018) have shown that D-L prior leads to posterior consistency and selection consistency.[5]

---

[5] Nevertheless, other global–local shrinkage priors can be used instead of D-L for the simple two-stage VB, depending on the researchers' preferences.

In both stages, each single equation to be estimated can be written in the following general form:

$$y = X\beta + e \qquad (3)$$

where $y$ is a $T \times 1$ vector of dependent variables, $X$ is a $T \times M$ matrix of explanatory variables, and $e$ is a $T \times 1$ vector of $iid$ error terms with $e_t \sim N(0, \sigma^2)$.

The hierarchical D-L priors for the $m^{th}$ element, for $m \in (1,...,M)$, of $\beta$ are as follows:

$$\begin{aligned} \beta_m \mid \phi_m, \tau &\sim DE(\phi_m \tau) \\ \phi_m &\sim Dir(a,...,a) \end{aligned} \qquad (4)$$

where $DE$ denotes the Double Exponential or Laplace distribution, and $Dir$ is the Dirichlet distribution. Finally, we set the following Gamma priors for $\tau$ and $\sigma^{-2}$:

$$\begin{aligned} \tau &\sim G(Ma, \tfrac{1}{2}) \\ \sigma^{-2} &\sim G(\underline{v}, \underline{s}) \end{aligned} \qquad (5)$$

The above hierarchical D-L priors for $\beta$ can be expressed as

$$\beta \sim N\left( \begin{bmatrix} 0 \\ 0 \\ \vdots \\ 0 \end{bmatrix}, \begin{bmatrix} \varphi_1 \phi_1^2 \tau^2 & 0 & \cdots & 0 \\ 0 & \varphi_2 \phi_2^2 \tau^2 & \cdots & 0 \\ \vdots & \vdots & \ddots & \vdots \\ 0 & 0 & \cdots & \varphi_M \phi_M^2 \tau^2 \end{bmatrix} \right) \qquad (6)$$

and

$$\varphi_m \sim Exp(\tfrac{1}{2}) \qquad (7)$$

Note that for the D-L priors, the only prior we need to select is $a$. This makes the prior sensitivity analyses a lot easier. With Normal prior for $\beta$ and Gamma prior for $\sigma^{-2}$, their posteriors can be drawn by Gibbs sampler or VB as described in Gefang et al (2020). To save space, we relegate the Gibbs sampler and VB algorithms to the [online appendix](#).

When the number of paramaters is large, Gibbs sampler can be time consuming. We therefore propose to incorporate VB, which is a fast alternative to Gibbs sampler, into the following simple two-stage algorithm to estimate model (2):

---

Simple two-stage VB Algorithm

---

1st Stage:

For $i = 1, ..., N$, run in parallel to predict $y_{it}$ using all the exogenous variables as instrumental variables

- Initialize the parameters and hyperparameters
- Run VB until the changes in paramters are negligible
- Save the predicted values of $y_{it}$

End

2nd Stage:

For $i = 1, ..., N$, substidude the endogenous variables by the predictive values derived in the first stage for each equation, run in parallel to estimate the $i^{th}$ row of $W^*$ and $(\theta_{i1}, ... \theta_{ik})'$,

- Initialize the parameters and hyperparameters
- Run VB until the changes in paramters are negligible
- Save the parameters

End

---

3. **Monte Carlo experiments**

To evaluate the performance of the simple two-stage VB, we simulate a large number of datasets using model (2) with combinations of various $N$, ranging from 30 to 500, and various $T$, ranging from 20 to 50. In all cases, we construct $W^*$, the spatial weights matrix, in three steps. First, we use the '$q$ ahead and $q$ behind' method of Kelejian and Prucha (1999) to establish the basic relationship matrix to specify that each unit is only spatially associated with the $q$ units in front of it and the $q$ units after it. Then, following Kelejian and Prucha (1999), we row standardise the basic relationship matrix so that the sum of each of its row is equal to 1. Second, we generate the spatial parameter $\rho_n^*$ associated with each unit by multiplying a random uniformly drawn parameter from the open interval $(\rho_a^*, \rho_b^*)$ and 1 or -1, depending on the outcome of a random draw from a Bernoulli distribution, in the same fashion as tossing a fair coin. Finally, we multiply each row of the basic spatial relationship matrix by its corresponding $\rho_n^*$ to get $W^*$. We generate the exogenous variables $x_{nt,r}$, for $r \in \{1,..k\}$, and the $iid$ disturbances from $N(0,1)$. The elements of the coefficients $\theta^*$ are drawn from the Uniform and/or Normal distributions.

Extensive Monte Carlo studies have shown that the simple two-stage VB works well for spatial panel models where $N \gg T$. To give some flavor, we next report the estimated results of three simulated datasets where $T = 20$, and $N = 30$, 300 and 500 respectively.

When generating the data, we set $q = 3$ so that each unit is spatially related to its 6 nearest neighbours. We then compute $\rho_n^*$, the spatial parameter for each unit, by multiplying a random draw from the Uniform distribution $U(0.6,,0.99)$ and 1 or $-1$, depeding on the outcome of a random draw from $Bernoulli(1/2)$. Finally, we generate two exogenous variables for each unit, and randomly draw their cofficients from $U(1,2)$ and $N(0,2)$, respectively.

Our simple two-stage VB algorithm is super fast. On a laptop[6], each stage takes around 2 seconds to estimate for panel data models where $N = 30$, around 2 minutes to estimate for models where $N = 300$, and around 10 minutes to estimate for models where $N = 500$.

### 3.1 The Spatial Weights Matrix

Figures 1(a), 2(a) and 3(a) plot the true spatial weights matrices we used to generate the simulated data.[7] The statistics of the non-zero elements of the true spatial weights matrices are reported in Table 1.

Table 1. Summary Statistics of the Off-diagonal Elements in the Spatial Weights Matrices

|  | N=30 | | N=300 | | N=500 | |
| --- | --- | --- | --- | --- | --- | --- |
|  | Mean | RMSE | Mean | RMSE | Mean | RMSE |
| Positive Elements | 0.1371 | 0.0176 | 0.1290 | 0.0170 | 0.1295 | 0.0187 |
| Negative Elements | -0.1246 | 0.0148 | -0.1264 | 0.0178 | -0.1283 | 0.0174 |

Figures 1(b), 2(b) and 3(b) plot the mean estimated spatial weights matrices. By visual inspection, we can find that when $N = 30$, Figure 1(b) largely resembles Figure 1(a). However, the estimated spatial weights matrices plotted in Figures 2(b) and 3(b), where the ratio $\frac{N}{T}$ is of a much larger magnitude, only look vaguely similar to their true conterparts in Figures 2(a) and 3(a). This is confirmed by the summary statistics of the off-diagonal elements of the estimated spatial weights matrices reported in Table 2. In Table 2, the bias and root mean square error (RMSE) for positive

---

[6] Processor: 11th Gen Intel(R) Core(TM) i5-1145G7 @ 2.60GHz, 2611 Mhz, 4 Core(s), 8 Logical Processor(s)
 RAM: 16 GB.

[7] For $N = 300$ and $N = 500$, we only plot their first $30 \times 30$ submatrices in order to make the figures legible.

element are calculated by taking the average of the bias and RMSE of all the positive elements. Likewise we get the averaged bias and RMSE for the negative and zero elements, respectively.

When $N$ gets larger, the magnitues of the estiamted values for the true non-zero elements globaly shrink towards zero, which is not ideal.. However, despite the fact that the estimated values for the true zero elements are rarely zero, they tend to be much closer to zero than the estimated values for the true non-zero elements.

Table 2. Summary Statistics of the Bias of the Off-diagonal Elements in the Estimated Spatial Weights Matrices

|  |  | Mean Bias | RMSE |
|---|---|---|---|
| N=30 | Positive Elements | 0.0737 | 0.1131 |
|  | Negative Elements | 0.0503 | 0.1067 |
|  | Zero Elements | 0.0045 | 0.1098 |
| N=300 | Positive Elements | 0.1216 | 0.0066 |
|  | Negative Elements | 0.1192 | 0.0065 |
|  | Zero Elements | 0.0002 | 0.0065 |
| N=500 | Positive Elements | 0.1249 | 0.0038 |
|  | Negative Elements | 0.1240 | 0.0038 |
|  | Zero Elements | 0.0001 | 0.0038 |

A natual question that arises is to what degree is the estimated spatial weights matrices like the true spatial weights matrix. To measure the similarities between the estimated spatial weights matrices and their corresponding true weights, we report their simple correlation and SSIM of Zhou et al. (2004) in Table 3.

Table 3. Similarities between the Estimated and True Spatial Weights Matrices

|  | corr2 | SSIM |
|---|---|---|
| N=30 | 0.9371 | 0.8002 |
| N=300 | 0.7407 | 0.9574 |
| N=500 | 0.7153 | 0.9736 |

Both simple correlation and SSIM are positive and in the range from .715 to .937 (correlation) and from .800 to .974 (SSIM). Consistent with the messages revealed in Figures 1(b), 2(b) and 3(b), the value of simple correlations between the estimated spatial weights matrix and the true spatial weights matrix decreases when *N* increases. However, the structural similarities between the estimated spatial weights matrix and its corresponding true spatial weights matrix measured by SSIM increases when *N* increases. Despite simple correlation results indicate that there are less similarities between the estimated matrix and the true one when N gets larger, SSIM show that taking out the impacts of noies, when N gets bigger, the estimated spatial weights matrix is actually becoming more similar to its true counterpart. This finding is particularly encouraging as it implies that we can rely on the estimated spatial weights matrix to make sensible inferences.

**3.2 Effects Estimates**

LeSage and Pace (2018) argue that effects estimates shall be the focus of Monte Carlo excersises because applied practitoners are more interested in the direct and indirect effects of changes in a explainatory variable on the dependent variable than any individual parameters. In the spirit of LeSage and Pace (2009), we define the effects matrix of model (2) as

$$\partial y / \partial X_r = (I_N - W^*)^{-1} diag(\theta_{ir},...,\theta_{Nr}),$$

where $r \in [1,...,k]$. The diagonal elements of the partial derivative matrix, which reflects the own partial derivative, $\partial y_i / \partial X_{i,r}$, measures the own-regional direct effect, and the off-diagonal elements of $\partial y_i / \partial X_{i,r}$, by contrast, measures the corresponding indirect effects.[8]

---

[8] In LeSage and Pace (2009), the direct effects are defined as the average of the diagonal elements and the indirect effects are defined as the average of the off-diagonal elements. By contrast, we focus on each individual element of the partial derivative matrix.

In Monte Carlo, we calculate $\partial y/\partial X_r$ for $r=1,2$. The findings from $\partial y/\partial X_1$ and $\partial y/\partial X_2$ are qualitatively the same. To save space, we focus on examining the former, and henceforth call it the effects matrix, which measures the spillover effects of a one unit increase in $X_1$. The summary statistics of the true effects matrices are presented in Table 4. In order that the positive and negative values do not cancel out each other, we focus on the absolute value of each element.

Table 4. Summary Statistics of the True Effects Matrix

|  | N=30 | | N=300 | | N=500 | |
| --- | --- | --- | --- | --- | --- | --- |
|  | Mean | RMSE | Mean | RMSE | Mean | RMSE |
| Direct Effects | 1.4712 | 0.2676 | 1.4952 | 0.2844 | 1.4961 | 0.2919 |
| Indirect Effects | 0.0459 | 0.0796 | 0.0045 | 0.0289 | 0.0027 | 0.0225 |

Since the direct effects are much larger than the indirect effects, we plot the diagonal elements of the true effects matrices in Figures 4(a), 5(a) and 6(a); we then plot the off-diagonal elements of the true effect matrices in Figures 4(b), 5(b) and 6(b). For ease of comparison, we plot the diagonal elements of the estimated effects matrices in Figures 4(c), 5(c) and 6(c), followed by the off-diagonal elements of the estimated effect matrices plotted in Figures 4(d), 5(d) and 6(d).

Two important messages emerge from Figures 4-6. First, regardless whether $N$ is large or small, the estimated direct effects remain very similar to the true direct effects. Second, the larger is $N$, the smaller is the magnitude of the indirect effects in comparison with the true indirect effects, however, the estimated indirect effects matrix and its true counterparts seem to share the same amount of similarities regardless of the value of $N$. The first finding is not surprising as $(I_N - W^*)^{-1} = I_N + W^* + (W^*)^2 + \cdots$. With the infinity norm of $W^*$ being less than 1, $I_N$ plays a dominant role in calculating the direct effects, and the impact of $W^*$ only takes place via the diagonal elements of $(W^*)^2$, $(W^*)^3$ and so on, which decrease rapidly. The second finding, however, is more interesting. To further investigate what happens, we report the similarities between the true and estimated effects matrices in Table 5. Indeed, the estimated and true direct effects matrices are almost equivalent in all cases with both simple correlation measures and SSIM approaching 1. Measured by simple correlation, the estimated and true indirect effects matrices are highly correlated, with corr2 exceeding 0.8 in all cases. It is striking to observe that corr2 measures for indirect effetcs only drop slightly when $N$ increases from 30 to 300, then to 500, even though the corr2 values between the corresponding estimated and true spatial weights matrices decrease more markedly. SSIM measures for indirect effects are relatively modest in comparison with those reported for the spatial weights matrices. Yet, SSIM tends to go up when $N$ increases.

Table 5. Simliarities between the Estimated and True indirect Effects Matrices

|       | Direct Effects | | Indirect Effects | |
|-------|--------|--------|--------|--------|
|       | corr2  | SSIM   | corr2  | SSIM   |
| N=30  | 0.9993 | 0.9978 | 0.8271 | 0.5380 |
| N=300 | 1.0000 | 1.0000 | 0.8092 | 0.5883 |
| N=500 | 1.0000 | 1.0000 | 0.7878 | 0.5744 |

It is also worthwhile to point out that both the negative and positive spillover effects are recovered without problems. The implication of this finding is profound: the unrestricted model we have proposed is able to capture spillover effects of different signs if that is what is in the data.

4. **Application to EU regional GVA Data**

Our dataset contains 202 EU NUTS2 regions listed in the Data Appendix. We use annual data spanning from 1999 to 2019, including the annual estimates of NUTS2 regional Gross Value Added (GVA), the number of scientists and engineers (sci), the number of working age population with lower education attainment (low_edu), the number of working age population with higher education attainment (high_edu), gross fixed capital (cap), total employment (emp) and total population (pop).[9]

We consider the following dynamic spatial panel model

$$y_t = W^* y_t + X_t \beta + \varepsilon_t \tag{8}$$

where $y_t$ is the annual GVA growth rates, $X_t$ contains the lagged GVA growth rates, the initial values of GVA, and the growth rates of sci, low_edu, high_edu, capital, employment and population. All the variables are normalised for each region. We use the simple two-stage VB approach to estimate the unrestricted spatial weights matrix $W^*$ and the parameter $\beta$. The coefficients of the pre-determined and exogenous variables are reported in Table 6.[10]

Table 6 Parameter Estimates

|      | Initial GVA | L GVA Growth | Scientists | Low_edu | High_edu | Capital | Emp | Population |
|------|-------------|--------------|------------|---------|----------|---------|--------|------------|
| Mean | **-0.1527** | **-0.1450**  | 0.0307     | 0.0256  | -0.0164  | **0.2093** | **0.2454** | 0.0394     |
| Std  | 0.0246      | 0.0242       | 0.0257     | 0.0240  | 0.0242   | 0.0236  | 0.0255 | 0.0238     |

The signs of the estimated parameters, except for those two associated with low_edu and high_edu, are in line with the findings of previous research: negative coefficients of Initial GVA and lagged GVA growth indicating GVA convergences among regions; positive coefficients of Scientist, Capital, Employment and Population in accord with economic theories. The coefficients of Low_edu and High_edu are not significant and of relatively small magnitudes; we shall refrain from overinterpreting them. But it would be interesting to investigate whether the unusual signs are caused by factors such as immigrations/emigrations and changes in high educations, in future research.

---

[9] The datasets are from ARDECO Database and data.europa.eu - The official portal for European data.

[10] We have tried different lag lengths and priors and found the results to be robust.

Now let us turn to the main concern of our empirical application, namely, examining the spatial interrelationships between regional GVA developments. We plot the histogram for the off-diagonal elements of the estimated spatial weights matrix in Figure 7. Clearly, a majority of the elements are of positive values and only a small number of them are negative, implying that there are more positive spillovers between regions than the negative ones.

Figure 8 plots the full estimated spatial weights matrix. It is evident that regions of the same country tend to share similar spatial interrelationship patterns. To highlight these patterns, we drop spatial elements whose absolute values are less than 0.003 from the spatial matrix and plot the remaining elements in Figure 9.

Five interesting messages emerge from Figures 8-9. First, a region tends to be more closely linked with regions within its own country than with other countries' regions. Along the zero-diagonal line, we can observe clear squared sub-matrices, associated with individual countries, whose elements are of much higher magnitudes than other elements either in the same rows or in the same columns. Second, the spatial linkages between German regions and regions of other countries are relatively loose. Only a small number of regions in Austria, Belgium, Czechia, France, Italy, Netherlands and Sweden are spatially linked with German regions. Third, in contrast to German regions which tend not to spatially interact with other countries' regions a lot, regions in Austria, Belgium, Spain, France, Italy and Portugal are all more closely spatially linked with each other. Fourth, regions in Greece are spatially linked with almost all regions of Spain and a small number of regions in Belgium, France, Italy and Portugal, but not much with regions of other countries. Fifth, a number of spatial elements are negative, a finding which we discuss below.

We plot the negative spatial elements in Figure 10. Five regions stand out: Pinzgau-Pongau (AT322) Salzburg und Umgebung (AT323) of Austria; Italy's Piemonte (ITC1) and Vest (RO42) of Romania. Those five regions have negative spatial relationships with almost all regions included in our sample. But the magnitudes of the spatial elements involving Germany and Sweden are much smaller than those of the elements involving other countries, suggesting that, if there are any crowding out effects, they do not affect Germany and Sweden much. If we consider crowding out effect of lesser magnitudes, all regions in Germany stand out. It is seen that increases in German regions' GVA growths will negatively impact the GVA growths in Greece. Although increases in Greek regional GVA growths will negatively affect German regional GVA growths, the impact here is small in comparison with German regions' impacts on Greek regions. Apart from Greek regions, we can see regions in Germany also exert some small negative impacts on regions in Romania and Poland. It seems possible that Germany is drawing in economic activities from Greece, and to a lesser extent, Romania and Poland.

The negative relationship between German and Greek GVA growth rates is consistent with the experiences of those countries from 2001, when Greece joined the euro area, until 2015, at which time the euro-area crisis is generally considered to have ended. The period 2001 to 2015 can be separated into two parts. From 2001 to 2008, German banks greatly increased their amount of lending into the Greek banking system to take advantage of the interest-rate differential in favour of Greek financial instruments and the absence of exchange rate risk. As a result, Greek GVA growth rose relative to German GVA growth. With the outbreak of the Greek sovereign debt crisis in 2009, the direction of capital flows between the two countries reversed as Greek residents, concerned with the stability of the Greek banking system, withdrew their savings from Greek banks and poured them into the safer German banks, reducing GVA growth in Greece and increasing GVA growth in Germany.[11]

Figures 11-14 plot the off-diagonal elements of $(I_N - W^*)^{-1}$, which govern the spillover estimates. For example, if we want to compute the spillover effects of an increase of the $i^{th}$ region's capital by 1, what we need to do is to multiply the $i^{th}$ column of $(I_N - W^*)^{-1}$ by 0.2393, the coefficient of Capital, the $j^{th}$ element of the product vector will be the amount of spillover from region $i$ to region $j$. Results presented in Figures 11-14 are consistent with those presented in Figures 8-10. To save space, we do not report findings that are qualitatively the same.

Our new findings indicate that the spatial interdependence between EU regional economies is more complicated than those documented in the existing literature. This is potentially of substantial importance, especially to policy makers in institutions such as the European Commission and the European Central Bank who design and implement regional policies in order to foster market integration and support economic growth.

5. **Conclusion**

If the a priori elicited spatial weights matrices are not sensible, inferences of the panel SAR models can be misleading or even worse than wrong. Although there are a number of techniques to test and compare models with different weights matrices, it is usually unrealistic to assume the spatial weights matrices selected by the tests results are the 'true' ones, because, while theories and conventions can only provide abstract guidance, the number of plausible spatial weights matrices can be unlimited. Hence it is important to let the data speak for themselves.

---

[11] For a detailed discussion, see Banerjee *et al.* (2021).

This paper extends Ahrens and Bhattacharjee (2015) and Gefang et al. (2023). Our main purpose is to uncover the spatial interdependence between cross-sectional units without imposing any a priori spatial weights matrices. We introduce a simple two-stage VB approach to estimating unrestricted panel SAR models where $N \gg T$. Using D-L prior, we are able to uncover the spatial linkage structure super-fast. Extensive Monte Carlo experiments show that our method works well. Keeping $T = 20$, we are able to identify the spatial weight matrices and spillovers that are highly correlated with their true counterparts for $N$ as large as 500.

In empirical applications, we investigate the spatial interdependencies of EU regional GVA growths. Apart from identifying prominent country clusters, we are also able to reveal a number of spatial relationships that have not been uncovered before, providing important information to support policy makers' decision.

**Acknowledgements:** This research used the ALICE High Performance Computing Facility at the University of Leicester

# Appendix. List of NUTS2 regions

| Austria | Germany | Greece | France | Hungary | Poland | Slovakia |
|---|---|---|---|---|---|---|
| AT11 | DE11 | EL30 | FR10 | HU21 | PL21 | SK01 |
| AT12 | DE12 | EL41 | FRB0 | HU22 | PL22 | SK02 |
| AT13 | DE13 | EL42 | FRC1 | HU23 | PL41 | SK03 |
| AT21 | DE14 | EL43 | FRC2 | HU31 | PL42 | SK04 |
| AT22 | DE21 | EL51 | FRD1 | HU32 | PL43 | |
| AT31 | DE22 | EL52 | FRD2 | HU33 | PL51 | |
| AT32 | DE23 | EL53 | FRE1 | **Italy** | PL52 | |
| AT33 | DE24 | EL54 | FRE2 | ITC1 | PL61 | |
| AT34 | DE25 | EL61 | FRF1 | ITC2 | PL62 | |
| **Belgium** | DE26 | EL62 | FRF2 | ITC3 | PL63 | |
| BE10 | DE27 | EL63 | FRF3 | ITC4 | PL71 | |
| BE21 | DE30 | EL64 | FRG0 | ITF1 | PL72 | |
| BE22 | DE40 | EL65 | FRH0 | ITF2 | PL81 | |
| BE23 | DE50 | **Spain** | FRI1 | ITF3 | PL82 | |
| BE24 | DE60 | ES11 | FRI2 | ITF4 | PL84 | |
| BE25 | DE71 | ES12 | FRI3 | ITF5 | **Malta** | |
| BE31 | DE72 | ES13 | FRJ1 | ITF6 | MT00 | |
| BE32 | DE73 | ES21 | FRJ2 | ITG1 | **Portugal** | |
| BE33 | DE80 | ES22 | FRK1 | ITG2 | PT11 | |
| BE34 | DE91 | ES23 | FRK2 | ITH1 | PT15 | |
| BE35 | DE92 | ES24 | FRL0 | ITH2 | PT16 | |
| **Cyprus** | DE93 | ES30 | **Netherlands** | ITH3 | PT17 | |
| CY00 | DE94 | ES41 | NL11 | ITH4 | PT18 | |
| **Czechia** | DEA1 | ES42 | NL12 | ITI1 | PT20 | |
| CZ01 | DEA2 | ES43 | NL13 | ITI2 | **Sweden** | |
| CZ02 | DEA3 | ES51 | NL21 | ITI4 | SE11 | |
| CZ03 | DEA4 | ES52 | NL22 | **Romania** | SE12 | |
| CZ04 | DEA5 | ES53 | NL23 | RO11 | SE21 | |
| CZ05 | DEB1 | ES61 | NL31 | RO12 | SE22 | |
| CZ06 | DEB2 | ES62 | NL32 | RO21 | SE23 | |
| CZ07 | DEB3 | ES63 | NL33 | RO22 | SE31 | |
| CZ08 | DEC0 | ES64 | NL34 | RO31 | SE32 | |
| **Estonia** | DED2 | ES70 | NL41 | RO32 | SE33 | |
| EE00 | DEE0 | **Finland** | NL42 | RO41 | | |
| **Luxembourg** | DEF0 | FI19 | **Latvia** | RO42 | | |
| LU00 | DEG0 | FI20 | LV00 | | | |

Footnote: The NUTS classifications can be found at https://ec.europa.eu/eurostat/web/nuts/background

Figure 1. Spatial Weights Matrices for $N = 30$

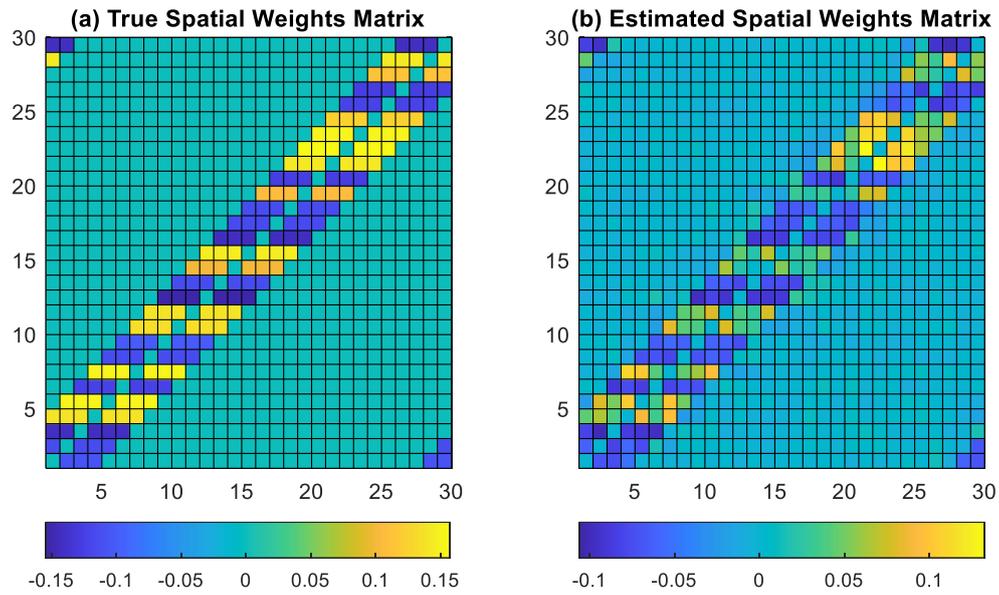

Figure 2. Spatial Weights Matrix for $N = 300$

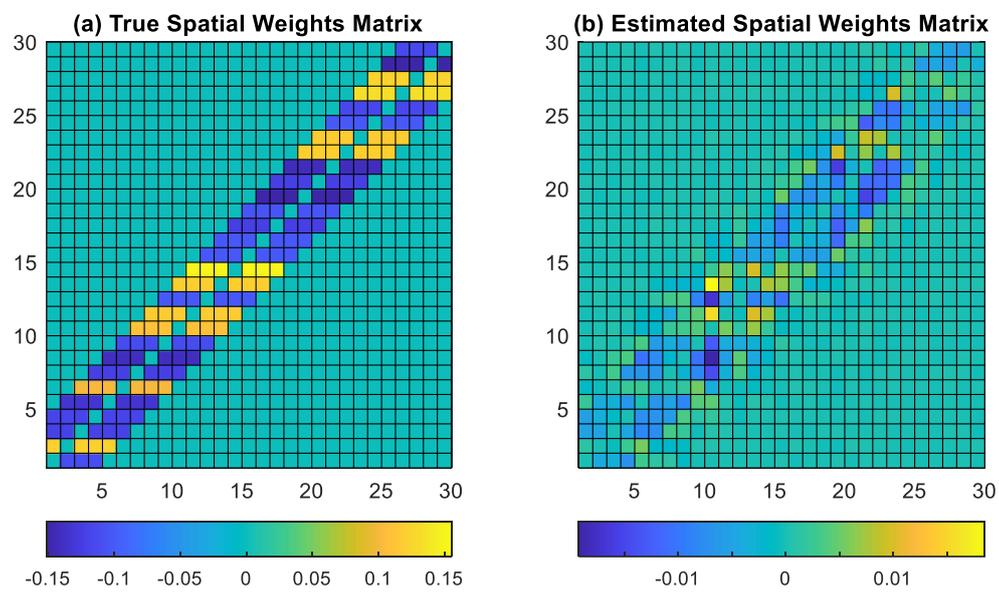

Figure 3 Spatial Weights Matrices for $N = 500$

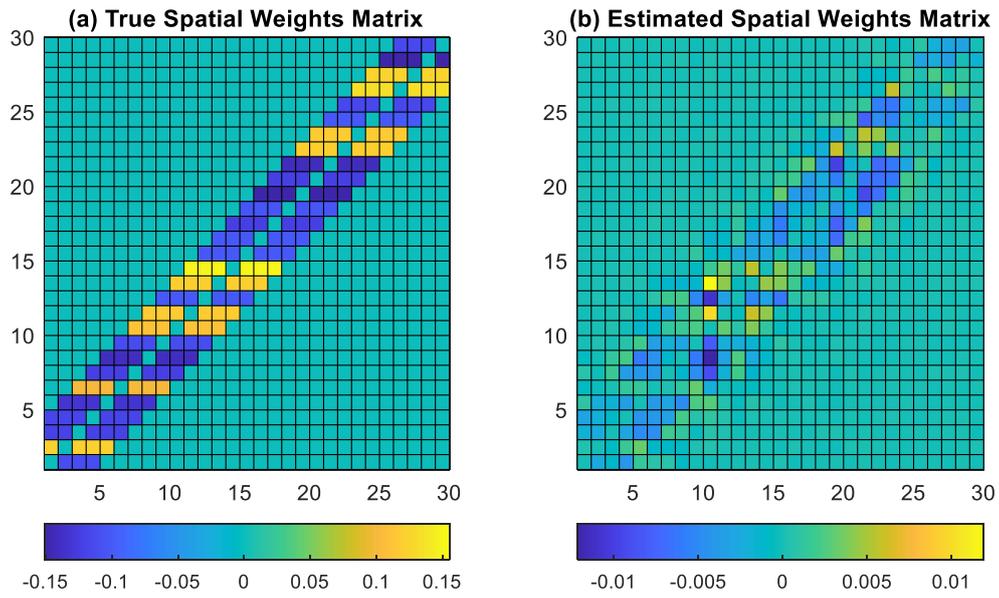

Figure 4 Effects Matrices for $N = 30$

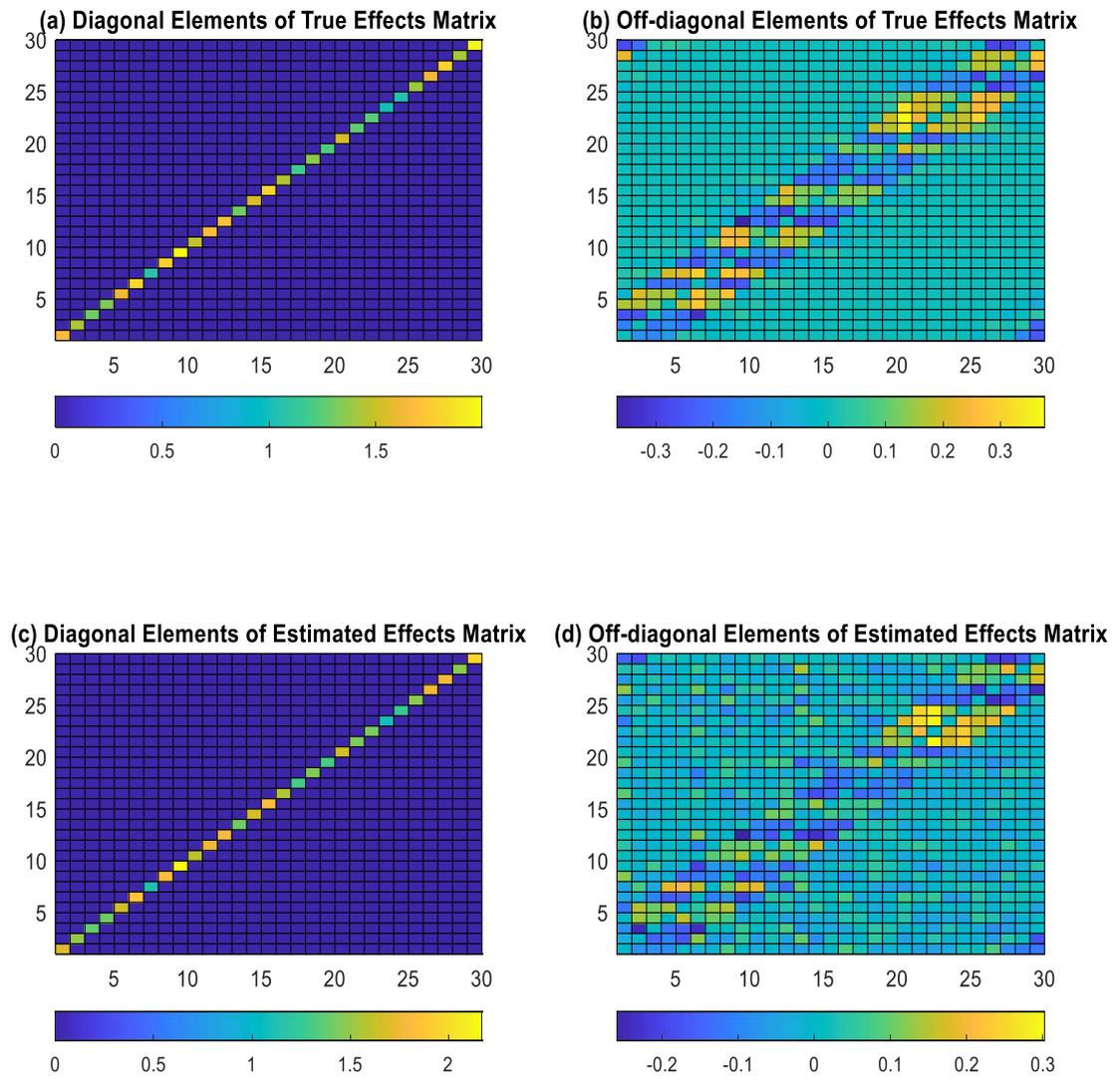

Figure 5 Effects Matrices for $N = 300$

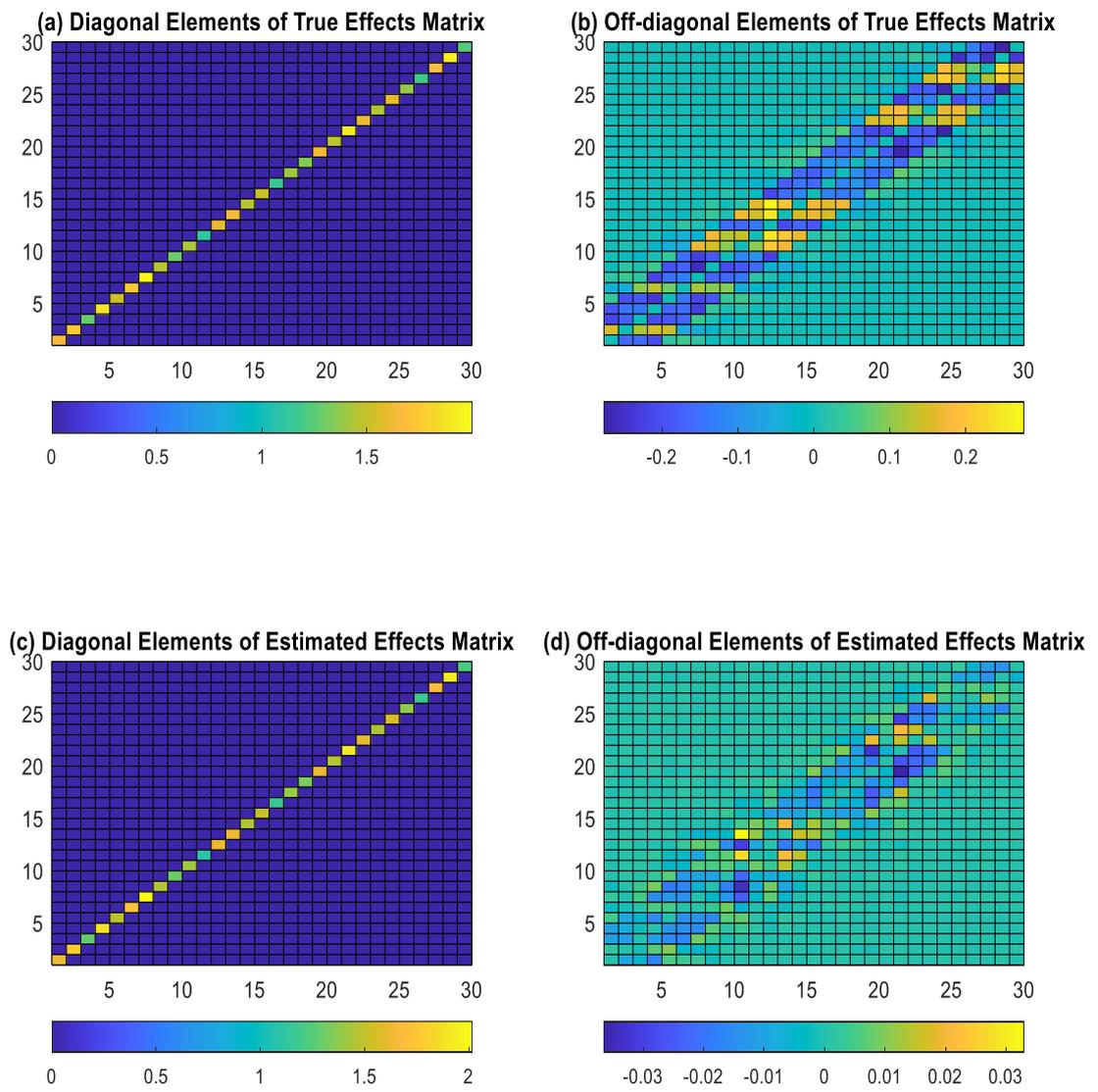

Figure 6. Effects Matrices for $N = 500$

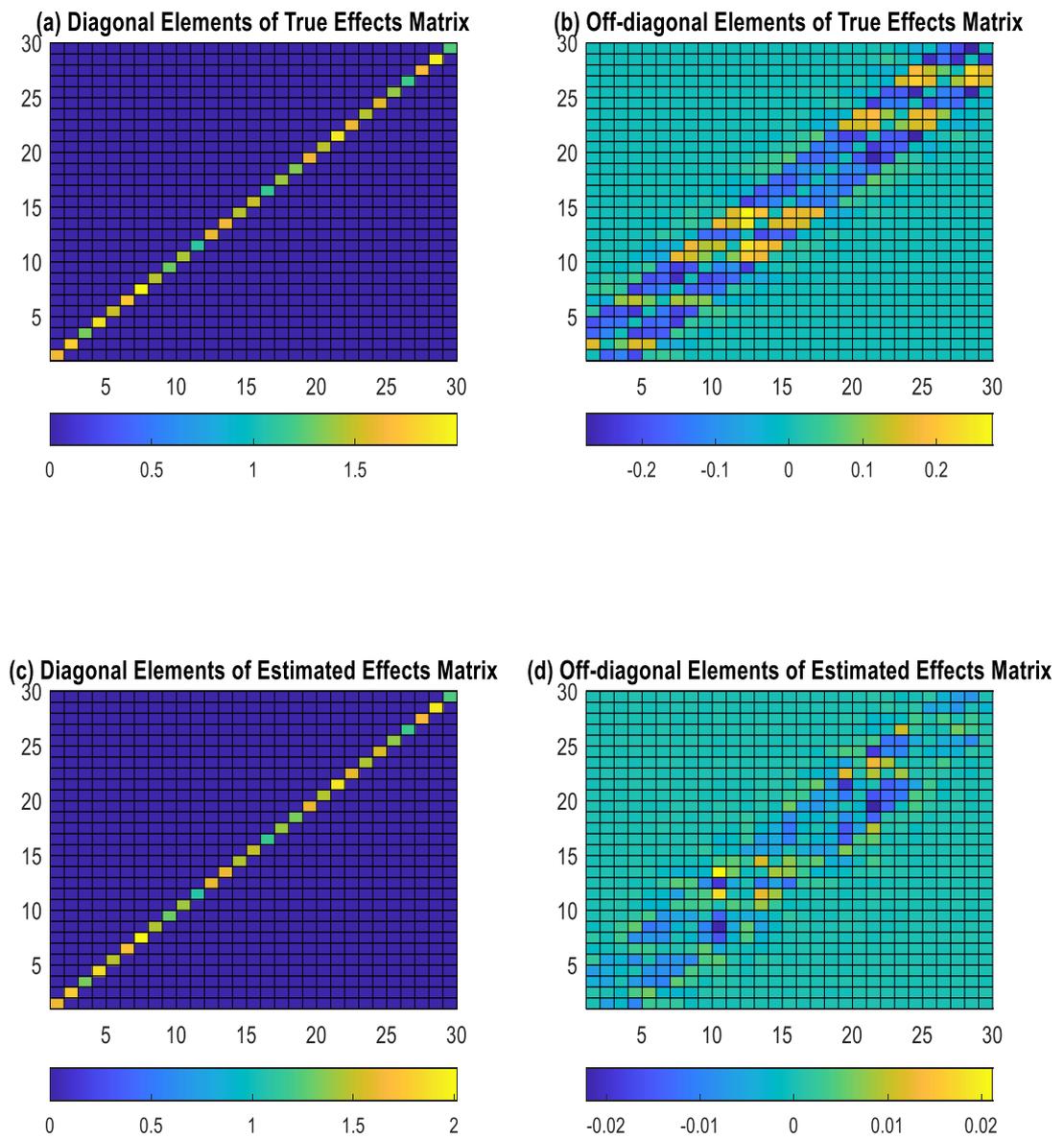

Figure 7. Histogram of the off-diagonal elements of the estimated spatial weights matrix for GVA data

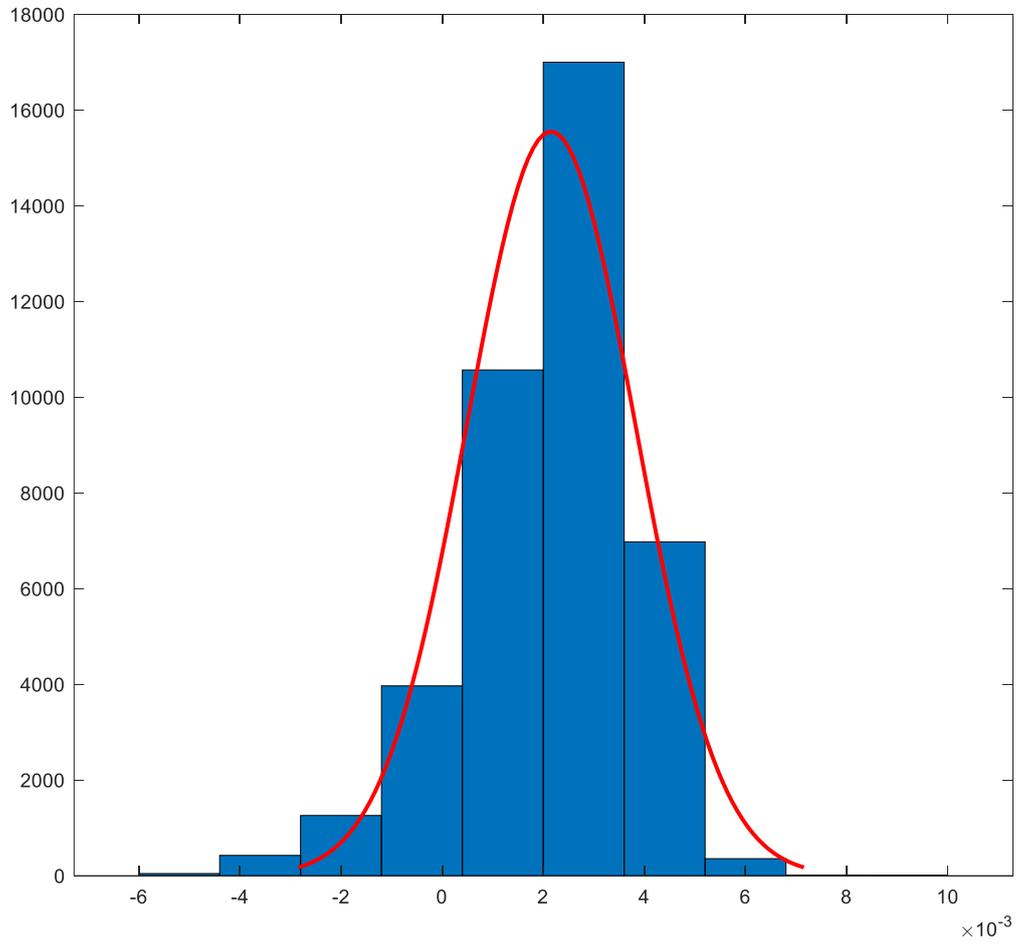

Figure 8. Estimated Spatial Weights Matrix for GVA Data (i)

Figure 9 Estimated Spatial Weights Matrix for GVA Data (ii)

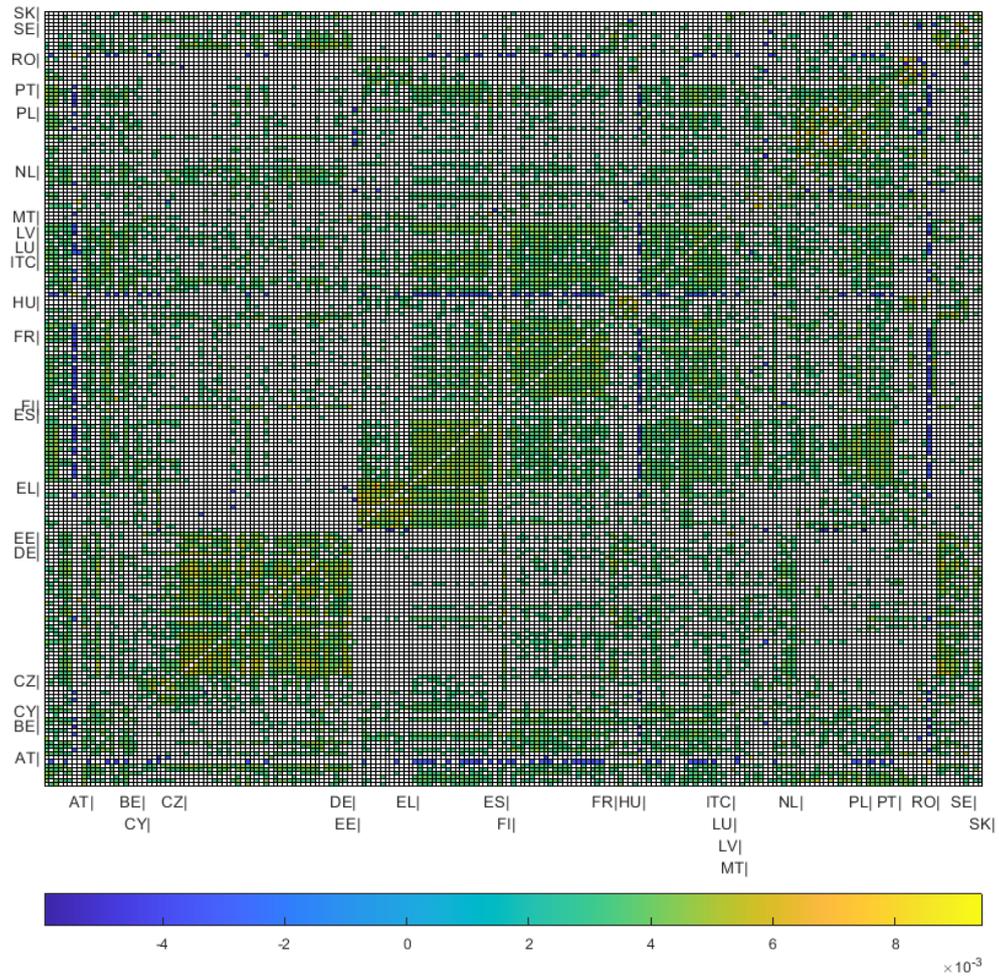

Figure 10 Estimated Spatial Weights Matrix for GVA Data (iii)

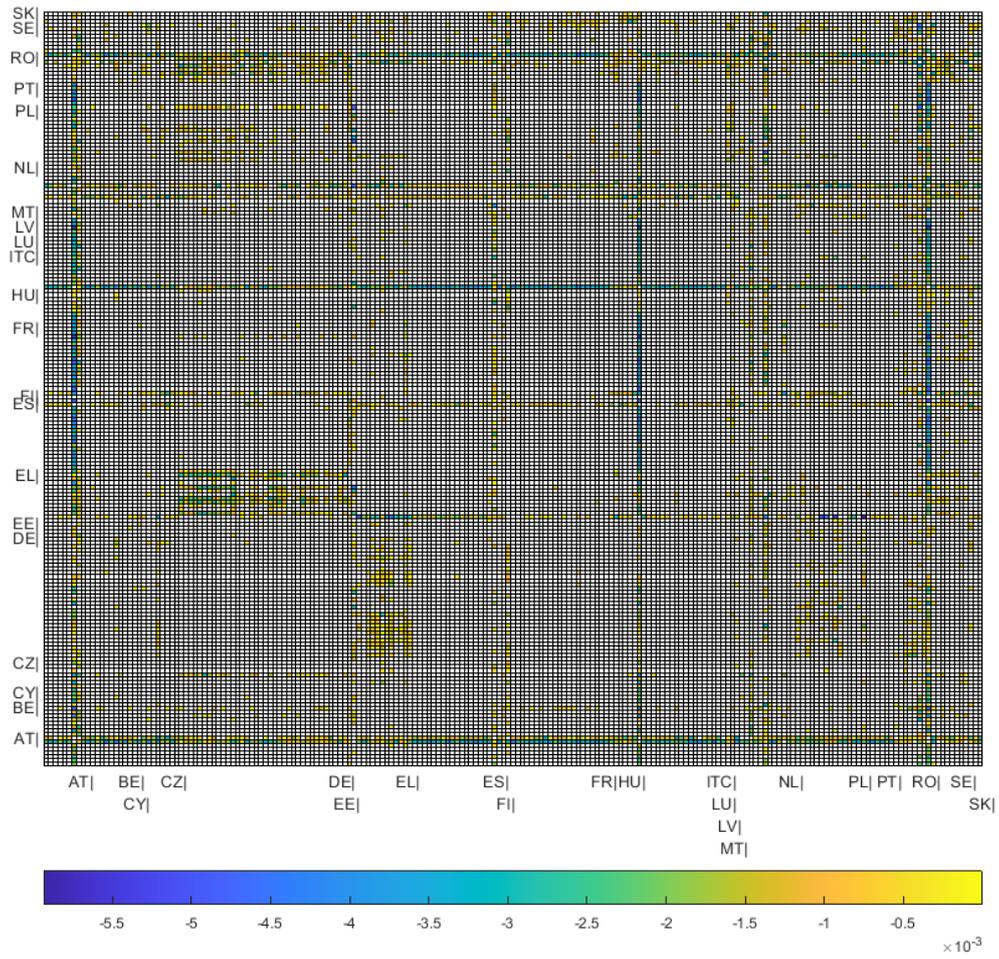

Figure 11. Histogram of the Off-diagonal Elements in $(I_N - W^*)^{-1}$

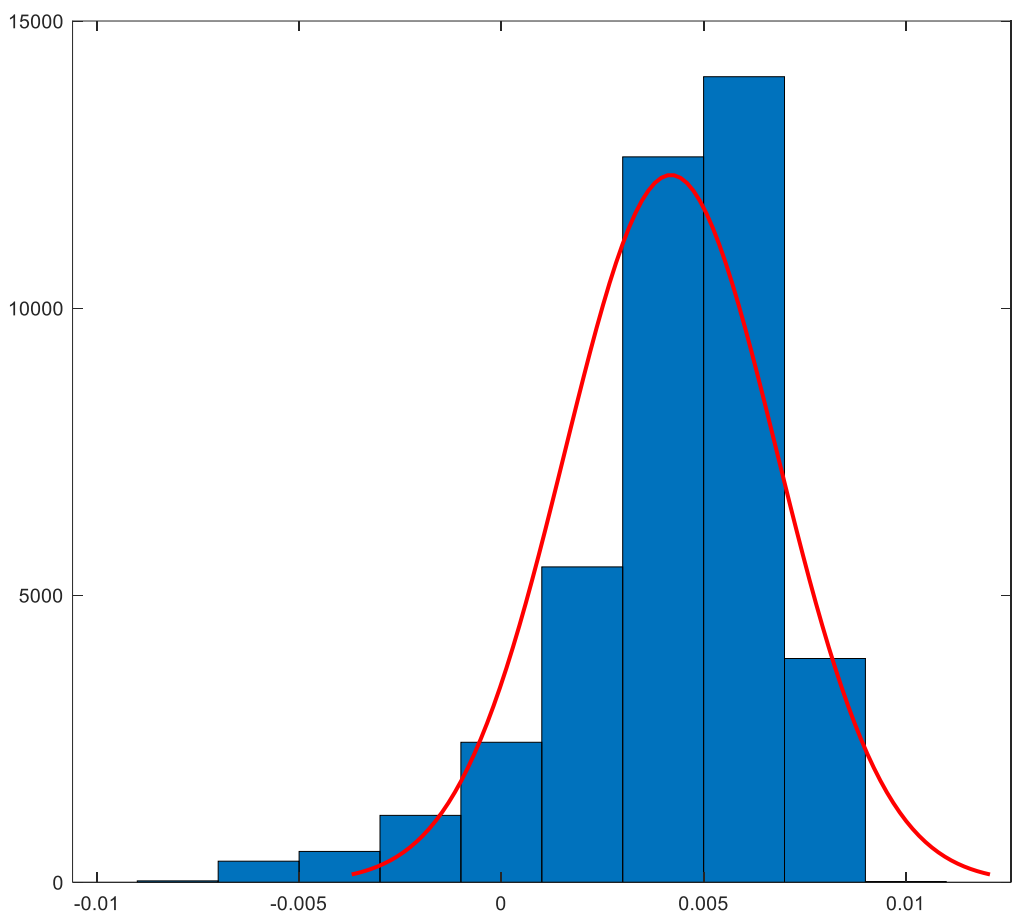

Figure 12 Matrix $(I_N - W^*)^{-1}$ with Zero Digonal Elemments (i)

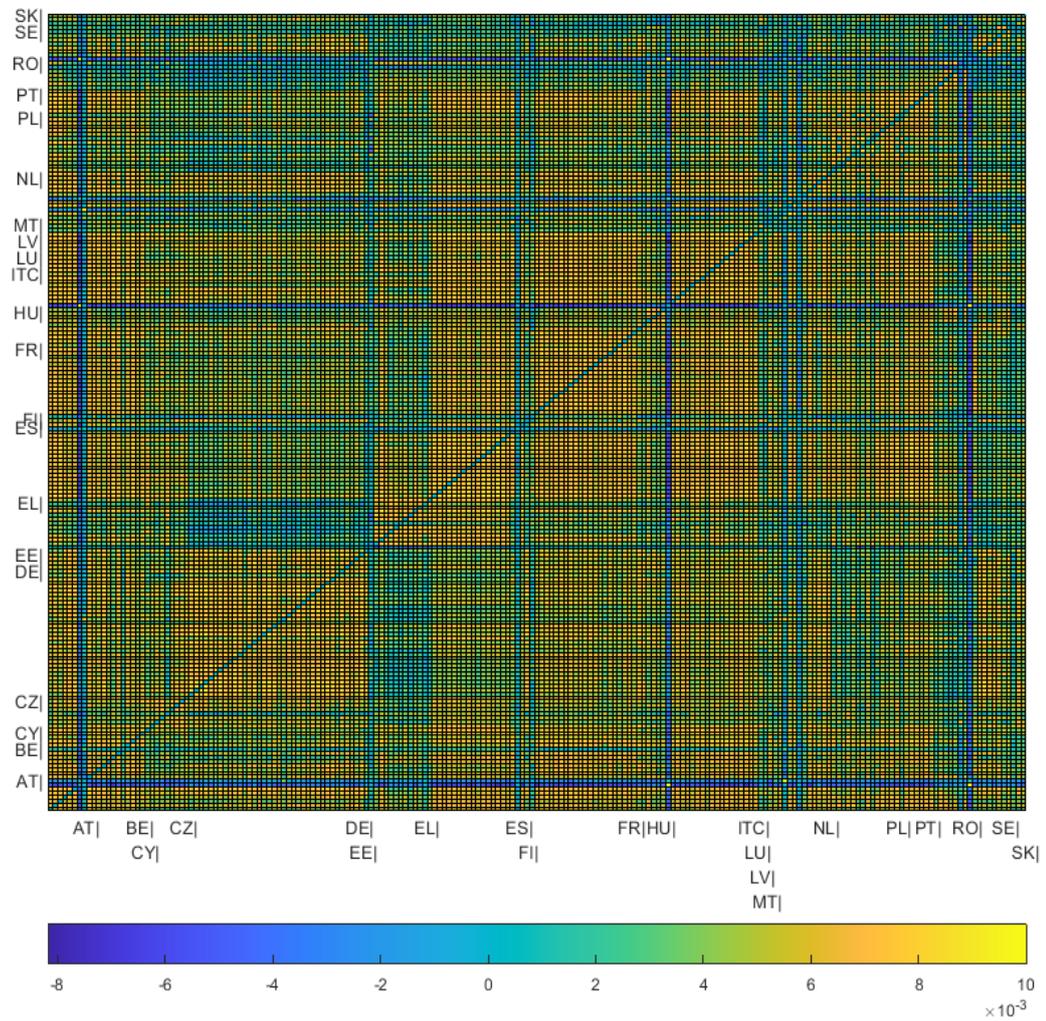

Figure 13  Matrix $(I_N - W^*)^{-1}$ with Zero Digonal Elemments (ii)

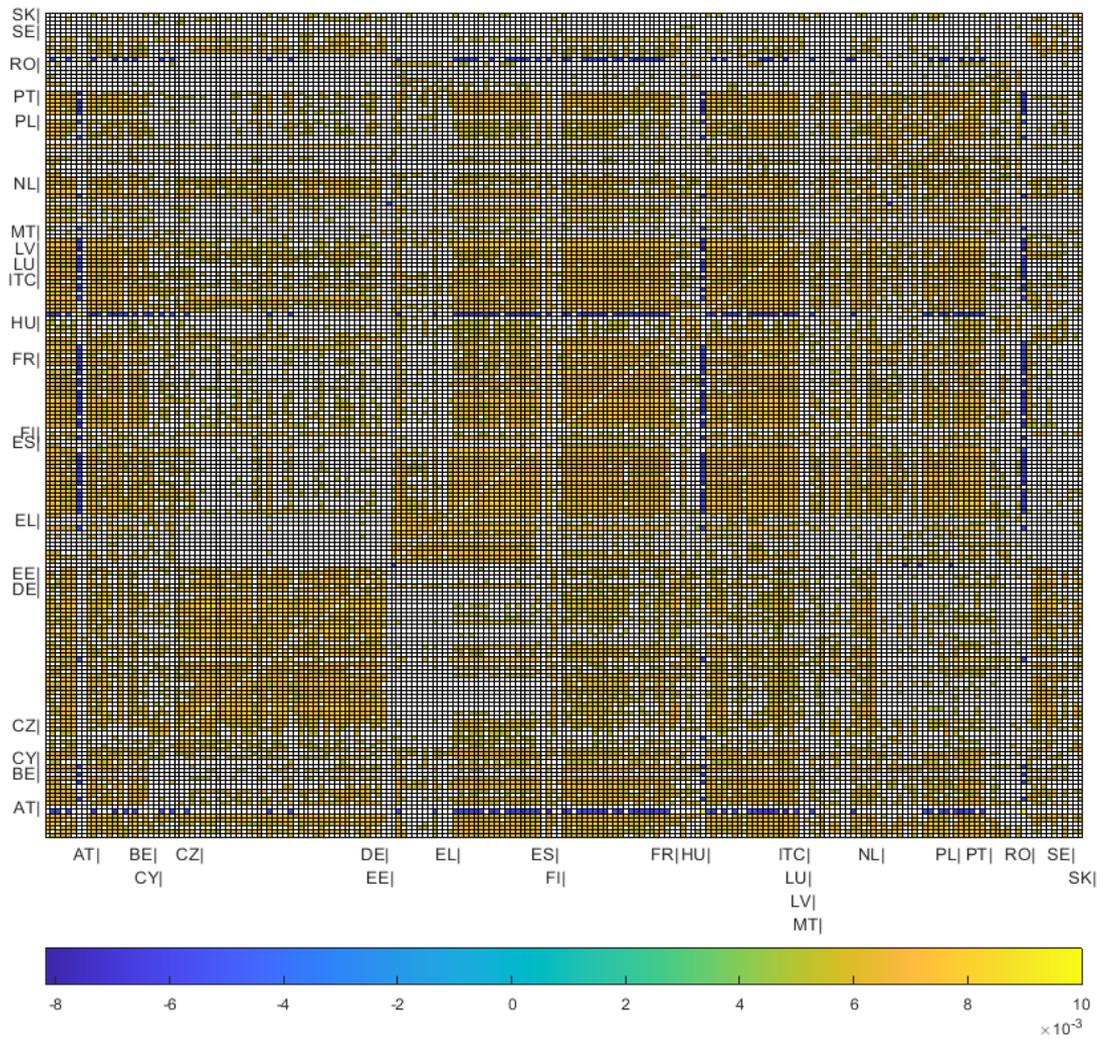

Figure 14 Matrix $(I_N - W^*)^{-1}$ with Zero Digonal Elemments (iii)

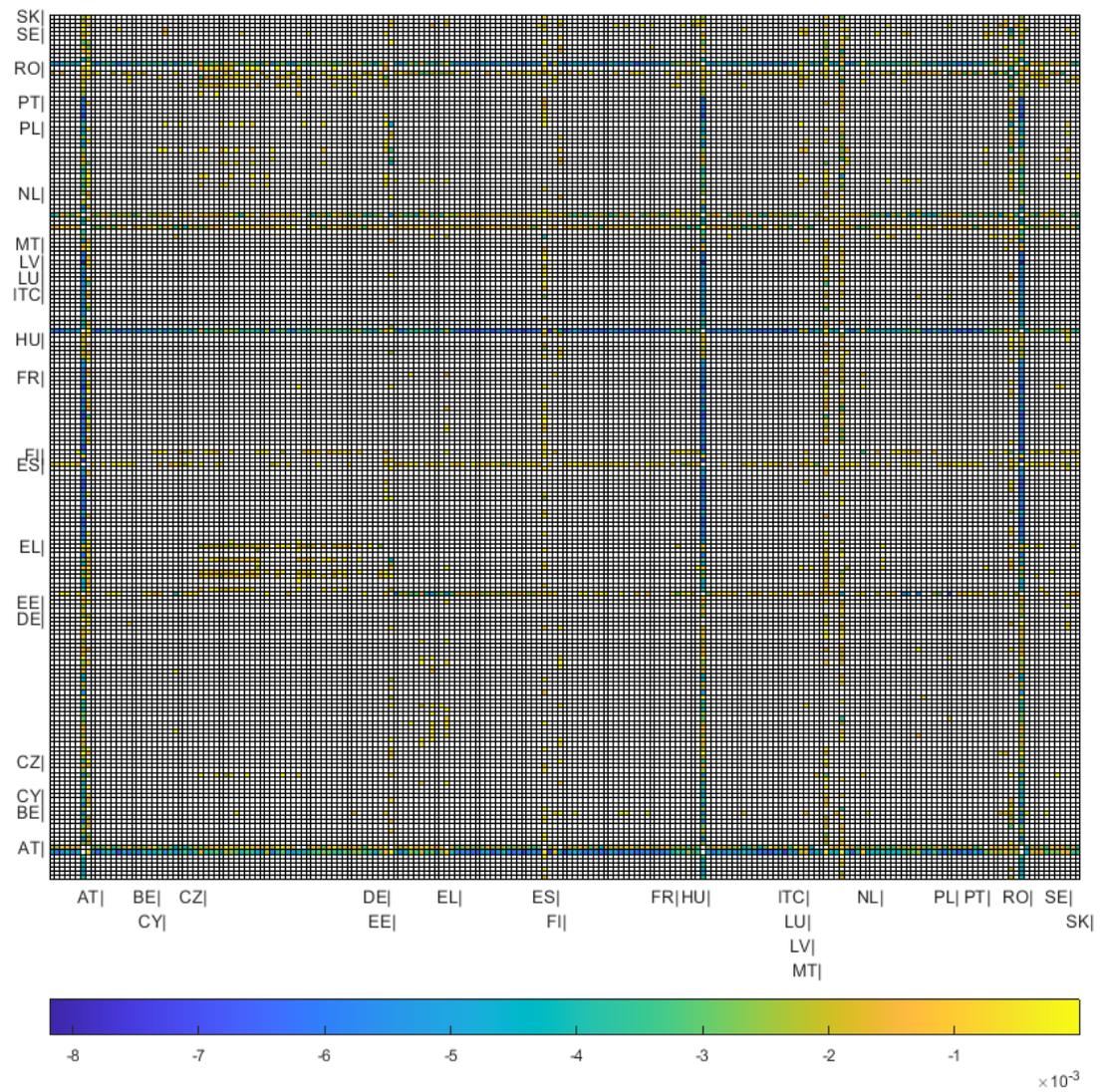